# A Practical Analysis Procedure on Generalizing Comparative Effectiveness in the Randomized Clinical Trial to the Real-world Trial-eligible Population

Kuan Jiang[a,b#], Xin-xing Lai[c,d#], Shu Yang[e], Ying Gao[c,d]*, Xiao-Hua Zhou[a,b]*

[a] *Department of Biostatistics, Beijing International Center for Mathematical Research, Peking University, Beijing, China PR;* [b] *Department of Biostatistics, School of Public Health, Peking University, Beijing, China PR;* [c] *Institute for Brain Disorders, Beijing University of Chinese Medicine, Beijing, China PR;* [d] *Department of Neurology, Dongzhimen Hospital, Beijing University of Chinese Medicine, Beijing, China PR;* [e] *Department of Statistics, North Carolina State University, North Carolina, USA*

[#] Kuan Jiang and Xin-Xing Lai contributed equally to this paper.

**\* Correspondence to**:
Ying Gao, M.D.
Institute for Brain Disorders, Beijing University of Chinese Medicine
No. 5 Haiyuncang, Beijing 100700, China PR
Email: gaoying97@163.com

Xiao-Hua Zhou, Ph.D.
Department of Biostatistics, School of Public Health, Peking University
Beijing 100871, China PR
Email: azhou@math.pku.edu.cn

**Author contributions**: Concept and design: Kuan Jiang and Xin-xing Lai; data collection: Xin-xing Lai; data analysis: Kuan Jiang; critical revision of the article for important intellectual content: Shu Yang and Xiao-Hua Zhou; study supervision: Xiao-Hua Zhou and Ying Gao. All the authors approved the final article.

**Financial Support**: This work was supported by the Beijing Nova Program of Science and Technology (Z211100002121061) and the Young Elite Scientist Sponsorship Program by the China Association for Science and Technology (2021-QNRC1-04).

**Declarations of interest:** none

**Word counts:** 4243

# A Practical Analysis Procedure on Generalizing Comparative Effectiveness in the Randomized Clinical Trial to the Real-world Trial-eligible Population

When evaluating the effectiveness of a drug, a Randomized Controlled Trial (RCT) is often considered the gold standard due to its perfect randomization. While RCT assures strong internal validity, its restricted external validity poses challenges in extending treatment effects to the broader real-world population due to possible heterogeneity in covariates. In this paper, we introduce a procedure to generalize the RCT findings to the real-world trial-eligible population based on the adaption of existing statistical methods. We utilized the augmented inversed probability of sampling weighting (AIPSW) estimator for the estimation and omitted variable bias framework to assess the robustness of the estimate against the assumption violation caused by potentially unmeasured confounders. We analyzed an RCT comparing the effectiveness of lowering hypertension between Songling Xuemaikang Capsule (SXC) —— a traditional Chinese medicine (TCM), and Losartan as an illustration. The generalization results indicated that although SXC is less effective in lowering blood pressure than Losartan on week 2, week 4, and week 6, there is no statistically significant difference among the trial-eligible population at week 8, and the generalization is robust against potential unmeasured confounders.

## Introduction

For evaluating drug efficacy, although RCT serves as the golden standard because of its perfect randomization, it lacks representativeness, leading to debates when applying results to real-world situations. As evidence, there are substantive studies indicating the inconsistency between results obtained from RCT and a real-world study due to different population heterogeneity(Hong et al. 2021; Lai et al. 2023; Munk et al. 2020; Rivera-Caravaca et al. 2018). Compared with RCT, a registry study provides a sample more representative of the general population than RCT, but unmeasured confounders could diminish its reliability(Victora, Habicht, and

Bryce 2004). Since RCT and real-world study have their strengths and weaknesses, it is important to generalize RCT results to a real-world trial-eligible population to draw correct conclusions(Sackett et al. 1996).

There are some statistical methods available to solve the problem, which have been proven reasonable mathematically. Intuitively speaking, we can model outcomes in RCT directly and make predictions in real-world data by the fitted model (the outcome-model-based estimator)(Colnet et al. 2023; Dahabreh et al. 2019; L. Nie and Soon 2010) or model the sampling score (probability of a unit being sampled into RCT population) and reweight RCT results by the inverse of sampling score (IPSW)(Buchanan et al. 2018; Cole and Stuart 2010; O'Muircheartaigh and Hedges 2014; Stuart et al. 2001; Tipton 2013). However, these methods need a correctly specified model to yield consistent estimation. The augmented inverse probability of sampling weighted (AIPSW) method models outcome and sampling score simultaneously and can get consistent estimation if either of the two models is correctly specified, also called doubly robust estimation(Dahabreh et al. 2019; Zhang et al. 2016). Recently, Lee et al. proposed a calibration weighting method to reweight units in the RCT sample (CW and ACW), and after calibration, the covariate distribution of the RCT sample empirically matches the real-world trial-eligible population(Lee et al. 2023). These methods generalize RCT results to a real-world trial-eligible population by the following data structure: treatment assignment in RCT, the RCT outcomes, and observed common covariates in both RCT and real-world data. For reviews on existing methods, readers can refer to them for formulas and relative proofs(Colnet et al. 2023; Dahabreh et al. 2019).

In addition to addressing observed covariates imbalance between RCT and real-world trial-eligible populations, exploring unobserved covariates is crucial and challenging for ensuring reliable generalization results by corresponding sensitivity analysis. Nguyen et al. (Nguyen et al. 2017; Nguyen et al. 2018)proposed a set of methods to evaluate the bias of IPSW and model-based estimators when unobserved confounders exist. Still, the methods rely on a strict parametric setting. Recently, based on Cinelli et

al.(Cinelli and Hazlett 2020),Chernozhukov et al.(Chernozhukov et al. 2023) and Douglas et al.(Faries et al. 2023), Huang proposed several tools to assess the robustness of generalization results, including numeric statistics (robustness value, minimum relative confounder strength, etc.), bias counterplots, and a formal benchmarking approach(Huang 2022). The advantage of Huang's method is that it does not require any assumptions on the data-generating process(Cinelli and Hazlett 2020; Faries et al. 2023; Huang 2022). Compared to other sensitivity analysis methods proposed by Nie et al.(X. Nie, Imbens, and Wager 2021), Colnet et al.(Colnet et al. 2022), it is more feasible in practical analysis.

So far, although there are many works investigating the generalization and sensitivity analysis methodology, few of them integrate them in comparative RCT and yield a reliable conclusion. In this article, we provide a statistical framework for generalizing findings in RCT to the real-world trial-eligible population based on existing statistical methods. Moreover, we utilized the framework of an RCT comparing the effectiveness of lowering blood pressure between two medicines as an example to illustrate the procedure. The article is organized as follows: Firstly, we will introduce the case study, including the background, datasets, summary statistics and our estimation target. Next, we will review relevant statistical methods to utilize in our practical analysis, including the AIPSW estimator and corresponding sensitivity analysis method based on the omitted variable bias framework briefly, and we will assemble the aforementioned methods in our framework to conduct a reliable RCT generalization project. Finally, we will use the framework to analyze the case as an illustration and showcase how to explain the results.

*Materials: Comparative RCT between SXC and Losartan and its generalization*

*Background*

Hypertension is one of the most prevalent chronic diseases globally. Losartan, a recognized medication, is often prescribed for treating hypertension(Kearney et al. 2005; Mills et al. 2016; Mills, Stefanescu, and He 2020). However, in China, Songling Xuemaikang Capsule (SXC) —— a traditional Chinese herbal medicine, is recommended to treat mild hypertension clinically as an alternative in certain official guidelines(Dongzhimen Hospital, Beijing University of Chinese Medicine and Editing Group of Guidelines for Prevention and Treatment of Hypertension 2023). Previous RCTs comparing the efficacy of lowering blood pressure between the two drugs demonstrated the non-inferiority of SXC to Losartan(Lai et al. 2022). However, as aforementioned, covariates distribution shift can make a difference when extending the conclusion to the real-world trial-eligible population. We aim to deal with the problem.

*Data Source, preprocessing and outcome*

We conducted two studies: an RCT and a registry study. The RCT was a multicenter, randomized, double-blind, non-inferiority trial that compared SXC with Losartan in terms of the efficacy of lowering blood pressure. It enrolled patients 18 to 65 years of age with mild essential hypertension (Grade I hypertension). In the treatment group, patients were assigned to receive SXC monotonously, and patients in the control group were assigned to receive Losartan. Details of the study design and the primary results were published previously(Lai et al. 2022). The registry study collected patients in the real-world practice setting who used SXC monotonously or a combination of SXC and other medicines to lower hypertension. We gathered baseline covariates and conducted follow-up visits in both studies at weeks 2, 4, 6, and 8.

In the RCT, the outcome was defined as an effect on lowering blood pressure, calculated by subtracting baseline systolic/diastolic blood pressure (BSBP/BDBP) from observed blood pressures at 2,4,6,8-week follow-up visits (denoted as DSBP/DDBP).

Efficacy differences between treatment and control groups were defined as the differences in means between the two groups. If the difference is greater than zero, then the effectiveness of SXC in lowering BP was worse than that of Losartan.

*Summary statistics of covariates*

We initially recruited 602 patients with grade I hypertension into the RCT, 300 of whom were assigned to the treatment group randomly. Meanwhile, 3,000 patients were enrolled in a registry study, and they only used SXC (N = 1567) or a mixture of SXC and other Western medicine (N = 1433) to lower blood pressure. After trimming the real-world data, dropping out units whose covariates are not supported by RCT data, 804 patients were left in the real-world cohort as a trial-eligible population.

We extracted seven common variables in both datasets: Age, Sex, Body Mass Index (BMI), Marriage, Smoking, baseline systolic blood pressure (BSBP), and baseline diastolic blood pressure (BDBP). Summary statistics of the variables are presented in Table 1. Of the baseline covariates, the mean of age was significantly greater in the real-world dataset (RWD) population than in the RCT group but was balanced between the treatment group and the control group within the RCT population. The standard deviation of BDBP in the RWD population is greater than the RCT population. Other baseline covariates were similarly distributed. Kolmogorov-Smirnov test (K-S test) was performed on continuous variables, and p-values were less than 0.05, indicating that the distribution of variables between the two populations was significantly different at the 0.05 level. Similarly, binary variables were tested by a two-sample Z-test. Box plots of continuous covariates were provided in Figure 1. In the real world, people using SXC were older than those in RCT and had lower BDBP than in the real-world trial-eligible population.

*Study Objective*

The primary objective of the study is to combine the RCT with real-world registry datasets to generalize RCT comparison results to a real-world trial-eligible population through the AIPSW estimator. Moreover, since the assumptions of generalization methods can be violated by unmeasured covariates, the robustness of the inference was assessed through corresponding sensitivity analysis.

*Methodology*

In this section, we will briefly review the methods that are utilized in the paper. We first introduce the notations and assumptions for the RCT generalization methods. Then, we will introduce the RCT generalization method and the corresponding sensitivity analysis method that we use in the article. Then, we provide a roadmap of adapting the methodology above into the comparative RCT, including generalization and corresponding sensitivity analysis.

*Notations*

We model the RCT data with sample size $n$ as a tuple of variable $(X_i, A_i, Y_i)$, $i = 1, \ldots, n$, where $X_i \in \mathcal{R}^p$ is a vector of $p$ covariates. $A$ is the treatment assignment and $A = 1$ denotes the treatment group. $Y$ is the observed outcome in the trial. In the real-world data of the trial-eligible population (RWD) with sample size $m$, we only collect the same covariates $X_i, i = n+1, \ldots, n+m$. Denote $S_i, i = 1, \ldots, n+m$ as the RCT membership, i.e. $S_i = 1$ if the unit is select into the RCT. Therefore, the two data structures can be presented as: RCT $(X_i, A_i, Y_i, S_i = 1), i = 1, \ldots, n$ and RWD $(X_i, S_i = 0), i = n+1, \ldots, n+m$.

According to the potential outcome framework in causal inference, we denote the potential outcome as $Y_i(a), a \in \{0,1\}$ and the observed outcome $Y_i = Y_i(a)$ if the unit is assigned $A_i = a$. We aim to estimate the average treatment effect (ATE) in the real-world trial-eligible population, i.e. $\mu = E[Y(1) - Y(0)]$. However, the ATE in RCT population is $E[Y(1) - Y(0)|S = 1]$, which does not equal to $\mu$ because of the selection bias.

*Assumptions*

To make the generalization, we need the following assumptions to make sure the consistency of the estimator.

- (**Consistency assumption**) $Y = Y(a), a \in \{0,1\}$. The assumption indicates that the observed outcomes of patients are one of the potential outcomes.
- (**Positivity of treatment probability in the trial**) $Pr[A = a|X, S = 1] > 0$ The assumption indicates that everyone in the RCT population can be assigned to a treatment group. The assumption holds in perfect RCT.
- (**Mean exchangeability in RCT**) $E[Y(a)|X, A, S = 1] = E[Y(a)|X, S = 1]$. The assumption indicates a perfect balance between the two groups in RCT. If perfect randomization is conducted in the design stage, the assumption holds.
- (**Positivity of trial participation probability**) $Pr[S = 1|X = x] > 0$, The assumption indicates patients in the real-world population have a positive probability of being selected into RCT. This assumption holds true for the trial-eligible population in the real world.
- (**Mean generalizability**) $E[Y|X] = E[Y|X, S]$, This assumption indicates that if sufficient variables are controlled, the outcome of a patient cannot be affected by their inclusion in the RCT group. The assumption is similar to the assumption of no unobserved confounders in causal inference and is not always straightforward. If unsure about controlling sufficient variables, sensitivity analysis can be conducted to assess the robustness of the results.

*Augmented Inversed Probability of Sampling Estimator*

The estimator has been proposed by Zhang et al(Zhang et al. 2016)., Darahreh et al(Dahabreh et al. 2019). and Colnet et al(Colnet et al. 2023). To utilize the estimator, we should fit two models: one is the model for the sampling probability and the other is the outcome model. Denote $\hat{\mu}$ as the AIPSW estimator for $\mu$, the formula is

$$\hat{\mu} = \frac{1}{n+m}\sum_{i=1}^{n+m}\left\{\frac{S_i A_i}{\hat{p}(X_i)\hat{e}(X_i)}\{Y_i - \hat{g}_1(X_i)\} - \frac{S_i(1-A_i)}{\hat{p}(X_i)(1-\hat{e}(X_i))}\{Y_i - \hat{g}_0(X_i)\} + \{\hat{g}_1(X_i) - \hat{g}_0(X_i)\}\right\},$$

where $\hat{p}(X)$ is an estimator for RCT selection probability $Pr[S=1|X]$, $\hat{e}(X)$ is the estimator for propensity score in RCT, i.e. $Pr[A=1|X, S=1]$. Here we use $\hat{g}_a(X), a \in \{0,1\}$, to model the potential outcome $Y(a)$. The estimator has double robust property; that is, either the RCT selection model or the outcome model is correctly specified can make sure a consistent estimation for $\mu$, i.e. $\hat{\mu} \xrightarrow{P} \mu$. In practice, to avoid the extreme value of $\hat{p}(X)(\hat{p}(X)$ is closed to 0 or 1), we always use a normalized form of the estimator, for details one can refer to Dahabreh et al(Dahabreh et al. 2019).

The variance and 95% confident interval of the estimator can be estimated by the bootstrap method.

*Sensitivity Analysis for AIPSW Estimator*

In the former section, we mentioned that the key assumption, the mean generalizability assumption can be violated if we do not collect all confounder covariates. Thus, the corresponding sensitivity analysis should be conducted to assess the robustness of the generalization results. We here adapted a feasible sensitivity analysis method proposed by Huang et al.(Huang 2022), which can be applied to the AIPSW generalization method. The idea is based on the sensitivity analysis framework proposed by Cinelli et al(Chernozhukov et al. 2023; Cinelli and Hazlett 2020), and can be summarized in the following steps.

Firstly, denote $w_i^*$ as the weight without unmeasured confounder (true weight), $w_i$ is the estimated weight (which is biased if unmeasured confounders exist) and $\varepsilon_i = w_i - w_i^*$ is the weight bias. Similarly, we denote $\xi_i$ as the difference between the true individual-level treatment effect and estimated treatment effect. Then the bias of the AIPSW estimator can be represented as:

$$Bias(\hat{\mu}) = \rho_{\varepsilon,\xi}\sqrt{var(w_i)\frac{R_\varepsilon^2}{1-R_\varepsilon^2}\sigma_\xi^2},$$

where $\rho_{\varepsilon,\xi}$ is the correlation relationship between $\varepsilon$ and $\xi$. $\sigma_\xi^2$ is the variance of the outcome error and the upper bound can be estimated by the data. Similarly, with the upper bound of $\sigma_\xi^2$, we can estimate the bounds for $\rho_{\epsilon,\xi}$ and vary the parameter in the range. For $R_\varepsilon^2$, is the ratio of variances between $\varepsilon_i$ and $w_i^*$, i.e. $R_\varepsilon^2 = \frac{var(\varepsilon)}{var(w_i^*)}$. According to the decomposition $var(w_i^*) = var(w_i) + var(\varepsilon)$, we can range $R_\varepsilon^2$ in [0,1). Therefore, with an estimated upper bound of $\sigma_\xi^2$, and a coordinate of sensitivity parameter $(R_\varepsilon^2, \rho_{\varepsilon,\xi})$, we can estimate a bias for the AIPSW estimation.

Next, we consider the confounding strength that unmeasured confounders have at least to induce a bias equal to the AIPSW estimation. Huang extended the Robustness value proposed in Cinelli et al.(Cinelli and Hazlett 2020) to measure how strong a confounder must be for the bias to equal $100 \times q\%$ of the estimated effect, that is

$$RV_q = \frac{1}{2}\left(\sqrt{b_q^2 + 4b_q} - b_q\right), \text{ where } b_q = \frac{q^2\widehat{\mu^2}}{\sigma_\xi^2 var(w_i)}$$

With a little abuse of notation, we use $RV$ to refer to the case with $q = 1$. If an unmeasured cofounder with the strength $\rho_{\varepsilon,\xi}^2 = R_\varepsilon^2 = RV$, then the AIPSW estimation can be totally offset by the confounding bias. In fact, a bias contour plot (x-axis: $R_\varepsilon^2$, y-axis: $\rho_{\varepsilon,\xi}^2$) can be presented for an intuitive illustration. The point which satisfies $\rho_{\varepsilon,\xi}^2 = R_\varepsilon^2 = RV$ is on the contour line with the bias equal to the estimation, and a stronger unmeasured confounder can even reverse the estimation. Intuitively, we call it a killer confounder.

Finally, we should tell the probability of the existence of such a killer confounder. The collected covariates can be utilized as benchmarking variables. We used MRCS (minimum relative confounding strength), $k_\sigma^{min}$ and $k_\rho^{min}$ to evaluate the relative strength a potential confounder needs to disturb the estimation. Intuitively, MRCS measures how much the relative confounding strength an omitted variable must have to result in a killer confounder, and $k_\sigma^{min}$ and $k_\rho^{min}$ can be similarly interpreted. We describe the benchmarking process on a bias contour plot, where the benchmark variables were annotated by their confounder strength. Here, for simplicity, we skip the formal benchmarking details, and readers can refer to Huang's work to learn further.

*Adaption in Comparative RCT Generalization*

Without loss of generality, we will take the comparative study between SXC and Losartan on lowering blood pressure as an example.

In the case study, we first use the AIPSW estimator to generalize the comparative result in the RCT to the real-world trial-eligible population and estimate the 95% confidence interval by the bootstrap method, denoted as $[\hat{\mu}_l, \hat{\mu}_u]$. We will conduct a sensitivity analysis on the bounds by the methods proposed above. For the benchmarking process, we assume that there is no such unmeasured confounder that has stronger confounding strength than the collected covariates. It is reasonable in clinical practice, because we always collect stronger confounders with priority. Therefore, if there is at least a benchmarking covariate that is not a killer confounder, then we consider the estimation to be robust.

There are three trivial scenarios:

- $\hat{\mu}_l < 0$ and $\hat{\mu}_u > 0$. The generalization result indicates that there is no statistically significant difference in efficacy between SXC and Losartan. Then, we conduct sensitivity analysis on $\hat{\mu}_l$ and $\hat{\mu}_u$. If no killer unmeasured confounders exist by the benchmarking variable process, we claim that the generalization result is robust.
- $\hat{\mu}_l > 0$. The generalization result indicates that the SXC is possibly less effective than Losartan on lowering blood pressure. Likewise, we can conduct sensitivity analysis on the lower bound $\hat{\mu}_l$. If $\hat{\mu}_l$ is robust, then we claim that SXC is less effective than Losartan in lowering blood pressure.

- $\hat{\mu}_u < 0$. It indicates that SXC is possibly more effective than Losartan in lowering blood pressure. We then assess the robustness of the upper bound $\hat{\mu}_u$, and if so, we claim the superiority of efficacy of SXC.

In fact, besides the special case above, we can list all scenarios and corresponding conclusions in the following table 2.

*Results*

In this section, we will implement the methods above on our SXC vs. Losartan datasets to illustrate the procedure. Firstly, we will compare the results between RCT and its generalization to real-world trial-eligible population. Besides, we also use the sensitivity analysis method above to assess the robustness of the generalization results at week 8 to investigate the long-term effect of SXC. The analysis on the RCT data has been published formerly(Lai et al. 2022), and we re-analysis it in the supplementary material.

*RCT generalization to RWD*

Comparison of the two drugs on lowering SBP and DBP in RCT and its generalization to real-world trial-eligible populations are in Figures 2(A) and 2(B). On lowering SBP, although no statistically significant result was reported in RCT, when generalized to the real world, the difference was significant except at week 8. In contrast, RCT and AIPSW estimations were similar in lowering DBP except at week 2, where AIPSW estimation was slightly significant. The comparison showed a discrepancy in conclusion between the two populations. In RCT, the efficacy of SXC on lowering SBP was comparable to Losartan, while in real-world trial-eligible populations, the conclusion held only in the long term (8w). As for DBP, conclusions from RCT and its generalization were similar, where SXC was not inferior to Losartan.

*Sensitivity analysis*

To compare long-term efficacy, we performed sensitivity analysis on the upper and lower bounds of the AIPSW estimation in week 8 as an example. In Figure 3, we make bias plots for the generalization bounds of DSBP (Figure 3(A)) and DDBP (Figure 3(B)). In Figure 3(A), for example, if unmeasured confounders exist in the green zone, then the upper bound can be reversed, indicating SXC's superiority over Losartan. In contrast, the lower bound can be reversed by unmeasured confounders in the red area, meaning that SXC is inferior to Losartan. The covariates serve as benchmarking variables, which are annotated in the figure based on their confounding strength. Under the former assumption that the unmeasured confounding strength should be restricted by the minimum confounding strength of observed covariates, we deduce that no such killer confounder exists and the generalization result is robust. The same logic is applied to Figure 3(B). Therefore, we conclude that there is no statistical difference in the efficacy between the two drugs, and the conclusion is robust.

Table 3 reported the minimum relative strength over benchmark variables for an unmeasured variable to be a killer confounder for the lower bound, which means that the confounder can overturn the negative value to zero or even positive if observed. Once the lower bound becomes positive, we can conclude SXC's significant inferiority to Losartan in lowering blood pressure, and then the generalization result on week 8 will be unreliable. Take DSBP as an example; the required relative strength (i.e. the ratio of bias caused by omitted confounders over benchmarking variables) for an unobserved confounder is no less than 58.16 times gender, 24.77 times age, and 23.2 times marriage, etc., to overturn the negative lower bound. Since we had collected all key confounders according to our clinical practice, we were confident in denying the existence of such unmeasured variables. A similar conclusion was made for DDBP at week 8.

*Discussion*

In this article, we proposed a framework of RCT results generalization to the real-world trial-eligible population based on combination of AIPSW estimator and omitted variable based framework. We used the framework to generalize the findings of a comparative RCT between SXC and Losartan and found inconsistencies between the RCT and its generalization results. Although the difference between the two drugs in treating hypertension was statistically insignificant in RCT, the efficacy of lowering SBP at week 2, week 4, and week 6 of SXC was inferior to Losartan significantly when it was generalized to the real-world trial-eligible population. As for lowering diastolic blood pressure, SXC was inferior to Losartan at week 4 but similar at week 6 and week 8. In the long term, we can conclude that SXC is not inferior to Losartan in treating hypertension in the real-world trial-eligible population. The difference in SXC efficacy in the two populations can be attributed to covariate distribution shifts, which is revealed in the summary description. Take age as an example. From Figure 1 and Table 1, the real-world population has more older patients than the RCT population. This may explain the difference in the study conclusions to some extent, perhaps because SXC works to lower blood pressure for seniors in the long term compared to Losartan. However, there was a higher proportion of older patients in the real-world trial-eligible population, thus, SXC behaved less effectively than RCT in the short term. In terms of generalization results, our analysis generated important hypotheses for further exploring the mechanisms of how SXC works and served as a guideline for using SXC to treat hypertension in the real-world population. In the real world, using a mixture of SXC and Losartan to treat hypertension may be a better solution to balance efficacy and side effects, which has been proven in existing studies(Yang et al. 2015).

The discrepancy between RCT and the real-world study was mainly due to covariates heterogeneity between RCT and real-world populations, including those that cannot be measured or observed directly. This has led to the inability of RCT populations to represent trial-eligible populations in real-world contexts accurately. There are numerous studies revealing the heterogeneity, many of which used matching methods to compare results obtained from RCT and real-world studies(Hong et al. 2021; Lai et al. 2023; Munk et al. 2020). For instance, a previous study utilized the propensity score matching (PSM) method to match SXC users in both RCT and real-world cohorts and concluded the SXC's non-inferiority in treating hypertension in the real-world population to the RCT population. In their framework, after balancing observed outcomes, the heterogeneity comes from external unmeasured covariates. However, to generalize RCT results to the real world, the imbalance between the observed variables of the RCT population and the real-world trial-eligible population must be considered. Therefore, in our study, we sought to simultaneously address heterogeneity from both observed and unobserved variables. We first used the AIPSW estimator to make a generalization, which addressed the covariates distribution shift between the two populations under study. However, the statistical method depends on the mean generalizability assumption, potentially threatened by unobserved variables. Therefore, we made a sensitivity analysis to illustrate that our generalization results were robust against heterogeneity from unobserved variables.

In addition to the AIPSW estimator, we applied other methods introduced in the previous section to generalize the RCT results. The corresponding estimation tables and figures can be found in supplemental materials. As shown in Figures S3.1 and S3.2, despite the similarity of point estimation, the three methods exhibited different standard errors. The difference in standard errors was explained by Dahabreh et.al., that when the probability model and outcome model are correctly specified, the large-sample variance of AIPSW estimators will be larger or equal to that of outcome model-based method and no larger than that of IPSW estimators(Dahabreh et al. 2019). The explanation was in accordance with our practical results.

The current combination (AIPSW + omitted variable bias-based sensitivity analysis) is not the only way that can be used to implement our framework, but it can be the best way. For example, if we use the IPSW estimator to generalize RCT, we can also use corresponding sensitivity analysis methods. We consider the advantages of the methods we used in our framework. Firstly, as discussed above, the AIPSW estimator is double robust, and its consistency can be less disturbed by model specification than the outcome-model-based estimator and IPSW estimator. Although the ACW estimator is also double robust, the corresponding sensitivity analysis method has yet to be developed. In addition, among existing sensitivity analysis methods, Huang's method is better because it does not require a parametric model and is easy to implement. Besides, the benchmarking process can utilize the covariates information and restrict the confounding strength of unmeasured confounders, which is more convenient to exploit in clinical practice.

Our study still has some limitations. Firstly, we relied on statistical generalization rather than evidence-based generalization, meaning we could not access real-world trial-eligible population data to verify our findings. Obtaining such evidence would be challenging, as designing RCT on the target population would be difficult, and conducting real-world studies may be affected by unmeasured confounders, which could introduce bias. However, the mathematical property of the estimator ensured that our results were reliable even without real-world evidence. Secondly, in the study, we dropped outpatients in the real-world cohort who did not meet the inclusion/exclusion criteria of the RCT to make a more accurate estimation. Therefore, our generalization conclusion was limited to the trial-eligible population. Recently, Paul et al. proposed a synthesis parametric model to deal with the non-positivity generalization problem. Still, this method requires subjective information, making it challenging to apply in practice(Zivich, Edwards, Shook-Sa, et al. 2023; Zivich, Edwards, Lofgren, et al. 2023).

*Acknowledgments*

We thank all patients and their families for participating in the registry or the SXC-BP trial and all the centers and investigators who devoted their effort and time to the two studies.

*Funding*

This work was supported by the Beijing Nova Program of Science and Technology under Grant [Z211100002121061]; and the Young Elite Scientist Sponsorship Program by the China Association for Science and Technology under Grant [2021-QNRC1-04].

*Disclosure statement*

No potential conflict of interest was reported by the author(s).

*Tables*

| Characteristics | RCT(N = 602) | | | RWD(N = 840) | p-value |
|---|---|---|---|---|---|
| | treatment group(N = 300) | control group(N = 302) | overall(N = 602) | | |
| Age | 50.3(8.83) | 50.3(9.39) | 50.3(9.11) | 53.1(8.31) | <0.001 |
| Sex | 0.537(0.499) | 0.517(0.501) | 0.527(0.50) | 0.556(0.497) | 0.29 |
| BMI | 25.1(3.38) | 24.9(3.02) | 25.0(3.20) | 24.3(3.20) | 0.005 |
| Smoking | 0.233(0.424) | 0.219(0.414) | 0.226(0.419) | 0.231(0.421) | 0.91 |
| Marriage | 0.977(0.151) | 0.964(0.188) | 0.970(0.170) | 0.989(0.103) | 0.01 |
| BSBP | 145.17(8.50) | 144.11(7.84) | 145.63(8.19) | 144.45(8.98) | <0.001 |
| BDBP | 92.01(5.03) | 92.01(5.55) | 92.0(5.29) | 86.88(7.58) | <0.001 |

Table 1 characteristic of baseline covariates and test between RCT and real-world data (RWD) population

| $\hat{\mu}_l$ | $\hat{\mu}_u$ | Conclusion |
|---|---|---|
| +,r | +,r | Inferiority |
| +,n | +,r | Inferiority/No difference |
| +,n | +,n | - |
| -,r | +,r | No difference |
| -,n | +,r | Inferiority/No difference |
| -,n | +,n | - |
| -,r | -,r | Superiority |
| -,r | -,n | Superiority/No difference |
| -,n | -,n | - |

Table 2. Summary of different sensitivity analysis scenarios and corresponding conclusions in the case comparing SXC and Losartan. The notation in the first two columns denotes the sign and robustness of the upper/lower bound. For example, for $\hat{\mu}_l$, '-,r' means that the estimation of lower bound is negative and it is robustness (no killer unmeasured confounders), while '+,n' is otherwise.

| Outcome | Variable | $k_\sigma^{min}$ | $k_\rho^{min}$ | MRCS | RV |
|---|---|---|---|---|---|
| 8W DSBP | gender | 15.11 | -2.96 | 58.16 | |
| | age | 0.66 | -6.17 | 24.77 | |
| | marriage | 7.41 | 1.69 | -23.2 | |
| | BMI | 1.67 | 1.04 | -6.77 | |
| | smoke | 6.04 | -0.97 | 12.08 | |
| | BSBP | 6.25 | -0.6 | 7.57 | |
| | BDBP | 0.08 | -2.88 | 2.9 | 0.01 |
| 8W DDBP | gender | 7.23 | -1.74 | 23.5 | |
| | age | 0.32 | 1.45 | -3.99 | |
| | marriage | 3.55 | 2.66 | -25.1 | |
| | BMI | 0.8 | 0.32 | -1.42 | |
| | smoke | 2.89 | 1.58 | -13.46 | |
| | BSBP | 2.99 | 1.07 | -9.29 | |
| | BDBP | 0.04 | 0.49 | -0.34 | 0.03 |

Table 3. Summary statistics and relative strength to benchmarking variables for AIPSW lower bounds at week 8

*Figures*

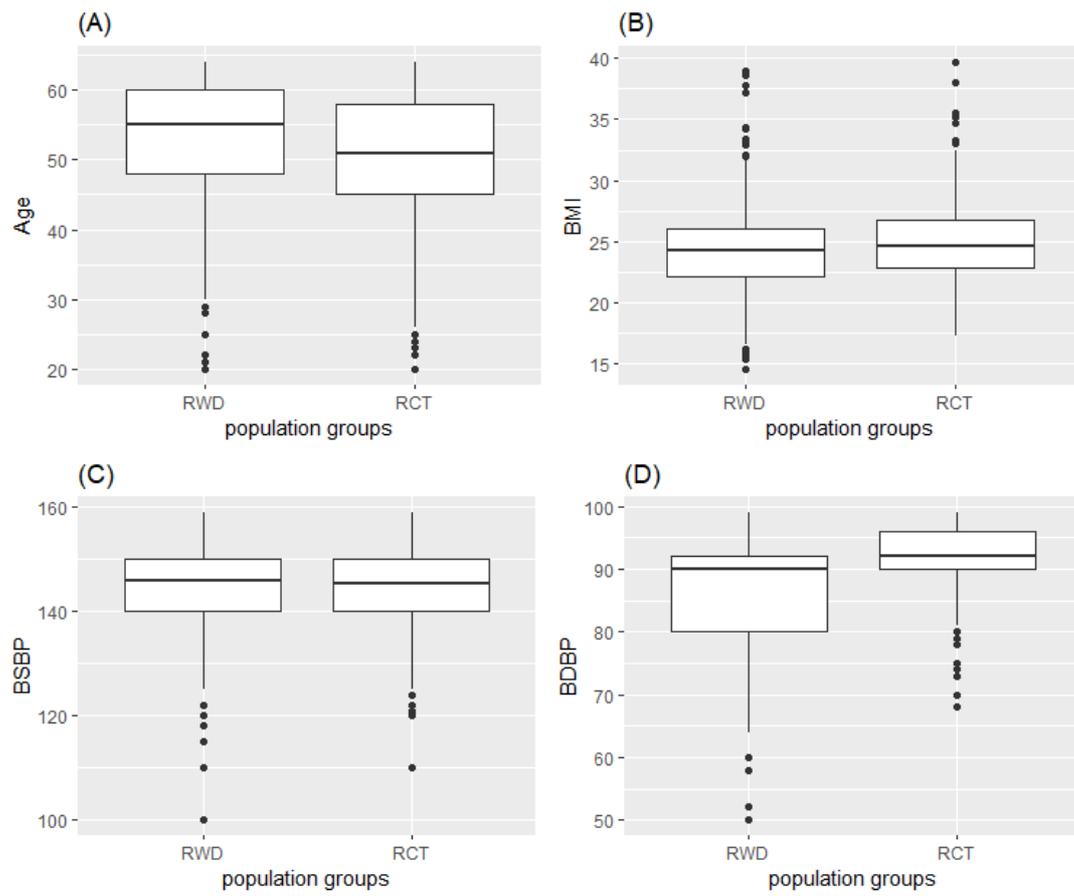

Figure 1. Box plots of continuous baseline variables in RWD and RCT sample: (A).Age, (B).BMI, (C). BSBP, and (D).BDBP

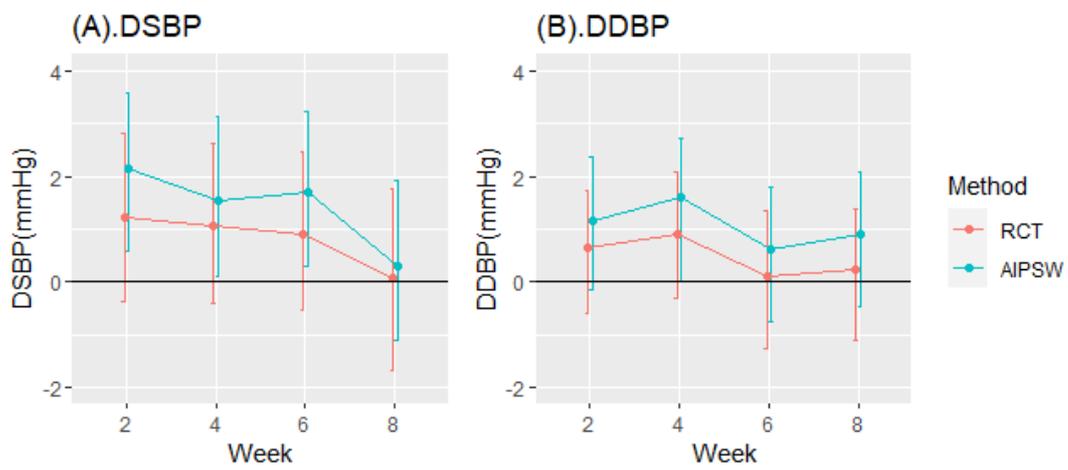

Figure 2. Difference of efficacy on lowering BP in RCT and its generalization to the real-world trial-eligible population by AIPSW estimator

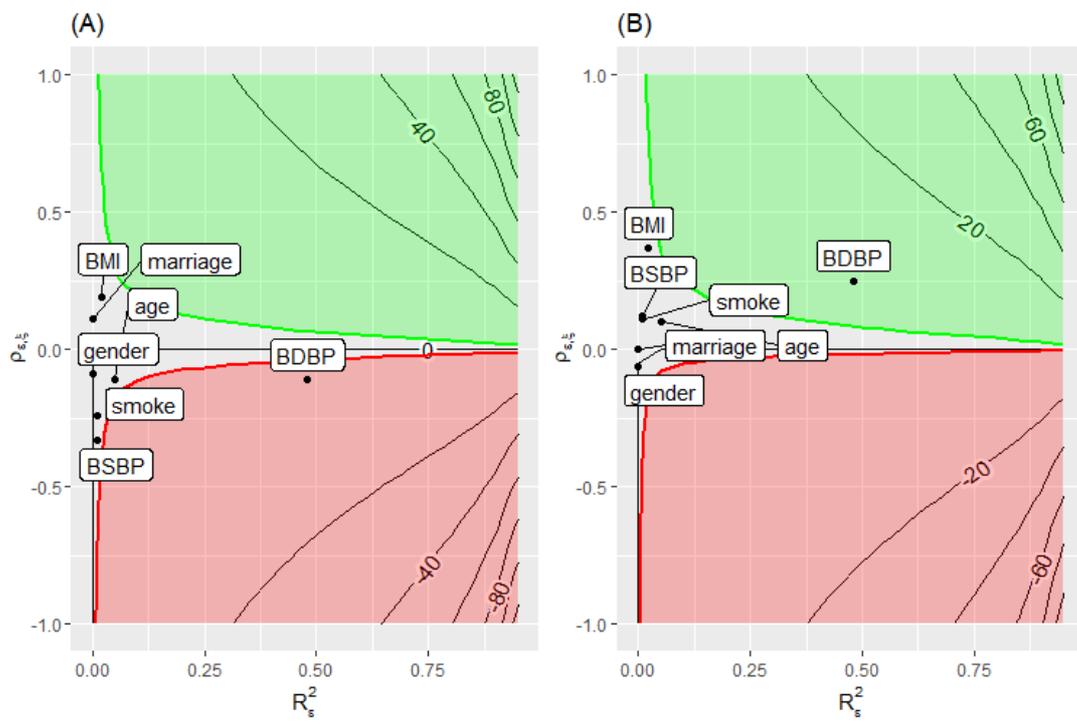

Figure 3. Bias plots for AIPSW bounds at week 8. Benchmarking variables are for the lower bound.